\documentclass{PoS}
\usepackage{amsmath,amssymb,amsthm}
\usepackage{amscd}

\newcommand{\CFock}{{C_{\mathrm{Fock}}}}
\newcommand{\DFock}{{\mathcal{D}'_{\mathrm{Fock}}}}
\newcommand{\HFock}{{H_{\mathrm{Fock}}}}
\newcommand{\RFock}{{R_{\mathrm{Fock}}}}

\newcommand{\calD}{{\mathcal{D}}}

\newcommand{\bbC}{{\mathbb{C}}}
\newcommand{\bbK}{{\mathbb{K}}}
\newcommand{\bbN}{{\mathbb{N}}}
\newcommand{\bbR}{{\mathbb{R}}}

\newcommand{\counit}{{\varepsilon}}

\newcommand{\potimes}{{\scriptstyle{\otimes}}}

\newcommand{\ix}[1]{{}_{\scriptscriptstyle(#1)}}

\newtheorem{prop}{Proposition}

\newtheorem{dfn}[prop]{Definition}

\title{Noncommutative version of Borcherds' approach to 
quantum field theory}

\ShortTitle{Noncommutative Borcherds' QFT}

\author{\speaker{Christian Brouder}\\
        Institut de Min\'eralogie, de Physique des Mat\'eriaux et
  de Cosmochimie, \\Sorbonne Universit\'es -- UPMC,
  Universit\'e Paris 6, UMR CNRS 7590, \\
  Mus\'eum National d'Histoire Naturelle, IRD UMR 206,
  4 place Jussieu,
  F-75005 Paris, France.\\
        E-mail: \email{christian.brouder@upmc.fr}}

\author{Nguyen Viet Dang\\
        Laboratoire Paul Painlev\'e, UMR CNRS 8524,\\
  Universit\'e de Lille 1, 59655 Villeneuve d'Ascq Cedex, France.}
\author{Alessandra Frabetti \\
 Institut Camille Jordan, Universit\'e Lyon1,
  UMR CNRS 5208,
\\B\^atiment Jean Braconnier,
 43 bd. du 11 novembre 1918,
   69622 Lyon, France.}

\abstract{Richard Borcherds proposed
an elegant geometric version of renormalized
perturbative quantum field theory in curved spacetimes,
where Lagrangians are sections of a Hopf algebra bundle over
a smooth manifold. 
However, this framework looses its geometric meaning
when Borcherds introduces a (graded) commutative normal product.
We present a fully geometric version of Borcherds' quantization
where the (external) tensor product plays the role of the normal product.
We construct a noncommutative many-body Hopf algebra and a module over
it which contains
all the terms of the perturbative expansion and we quantize
it to recover the expectation values of standard
quantum field theory when the Hopf algebra fiber is (graded)
cocommutative. This construction enables to the second quantize 
any theory described by a cocommutative Hopf algebra bundle.
}

\FullConference{Frontiers of Fundamental Physics 14 - FFP14,\\
                15-18 July 2014\\
                Aix Marseille University (AMU) Saint-Charles Campus, Marseille }
		
\begin{document}

\section{Introduction}
In an article entitled
``Renormalization and quantum field theory''~\cite{Borcherds-10},
Richard Borcherds described a rigorous approach to
renormalized perturbative quantum field theory in curved
spacetimes. Borcherds' approach is closely related to the causal
algebraic formalism~\cite{Brunetti2}, and it employs 
sheaf theory and Hopf algebras to 
achieve a particularly elegant and compact picture of 
quantum field theory (QFT).  In particular, the combinatorial aspects 
of quantization and renormalization
are completely taken care of by a Hopf algebraic structure.
Moreover, Borcherds' approach has definite advantages when it comes 
to generalization.
For example, the use of Hopf algebras is particularly powerful
to deal with systems involving an initial state which is not
quasi-free~\cite{BFP} and many of its tools
(for example vector bundles and Hopf algebras)
have natural noncommutative analogues that can be used to investigate
noncommutative versions of quantum field theory.

In the present paper, which is a sketch of a more
detailed article in preparation, we extend parts of Borcherds' approach
by replacing his graded commutative normal product of classical
fields by a tensor product which (i) allows us to
formulate a fully geometric version of second quantization,
(ii) provides a manageable topology for the 
many-body algebra, (iii) enables us to second quantize any
cocommutative Hopf algebra bundle.

\section{Hopf algebra bundles}
In this section we introduce some concepts that are
used in Borcherds' approach to QFT.
Classical fields are sections of vector bundles over the
space-time manifold $M$.
We first reformulate Borcherds' sheaves into 
more familiar sections of vector bundles.

Let $M$ be a smooth manifold and 
$F\overset{\pi}{\rightarrow} M$ a smooth vector bundle
over $M$~\cite{LeeDiff}. 
We denote by 
$\phi_\alpha: \pi^{-1}(U_\alpha)\to U_\alpha\times V$
the local trivializations (where $V$ is a vector space) and by 
$t_{\alpha\beta}$ the transition functions such that
$\phi_\alpha\circ\phi_\beta^{-1}(x,v)=(x,t_{\alpha\beta}(x)v)$,
where $\phi_\alpha\circ\phi_\beta^{-1}: 
(U_\alpha\cap U_\beta) \times V \to
(U_\alpha\cap U_\beta) \times V$ and where the isomorphism
$t_{\alpha\beta}(x)$
is an element of $GL(V)$. 
A vector bundle is an \emph{algebra bundle} if
the fiber model $V$ is an algebra over $\bbK$
(where $\bbK$ is $\bbR$ or $\bbC$) and if the
transition functions are algebra isomorphisms:
$t_{\alpha\beta}(x)(u \cdot v)=
t_{\alpha\beta}(x)(u) \cdot t_{\alpha\beta}(x)(v)$.
An algebra bundle is a \emph{Hopf algebra bundle} if
$V$ is a Hopf algebra over $\bbK$ and the transition
functions are Hopf algebra morphisms.
In particular, the coproduct sends $V$ to
$V\otimes V$, which is the fiber of the
(internal) tensor product of
vector bundles $F\otimes F\overset{\pi'}{\rightarrow} M$~\cite{LeeDiff}.

The space of sections $\Gamma(M,F)$ is an infinite-dimensional
vector space, but it is also a module over the ring $C^\infty(M)$
of $\bbK$-valued smooth functions: as such, it admits a (locally)
finite basis which allows to use simple linear algebra tools.
If $F$ is an algebra bundle, then the space of sections
$\Gamma(M,F)$ is an algebra over the ring $C^\infty(M)$:
if $\sigma_1$ and $\sigma_2$ are such sections with
$\phi_\alpha\big(\sigma_1(x)\big)=(x,v_1)$ and
$\phi_\alpha\big(\sigma_2(x)\big)=(x,v_2)$, then
$\phi_\alpha\big(\sigma_1\cdot\sigma_2 (x)\big)=(x,v_1\cdot v_2)$.
Similarly, the space of sections of a Hopf algebra
bundle is a Hopf algebra over the ring $C^\infty(M)$.
In particular, the coproduct is now a map
from $\Gamma(M,F)$ to
$\Gamma(M,F\otimes F)\cong\Gamma(M,F)\hat\otimes_{C^\infty(M)}
\Gamma(M,F)$.

Borcherds starts from a vector bundle
$E\overset{\pi}{\rightarrow} M$ of finite rank
whose sections
are the classical fields of the model. To define Lagrangian densities
as polynomials in the field and its derivatives, he
considers the infinite jet bundle 
$JE\overset{\pi}{\rightarrow} M$ 
and the Hopf algebra bundle $S(JE^*)
\overset{\pi}{\rightarrow} M$, which describes
the polynomial functions on $JE$.

For example, the element $L=
f+g_\mu+h_{\mu\nu}+k$, where
$f\in \Gamma(M,E^*)$, $g^\mu\in 
\Gamma(M,J^1E^*)$, $h^{\mu\nu}\in \Gamma(M,S^2(J^1E^*))$
and $k\in \Gamma(M,S^4(E^*))$, 
corresponds to the Lagrangian
density $L(\varphi)=\langle f,\varphi\rangle +
\langle  g^\mu ,\varphi_\mu\rangle
+ \langle h^{\mu\nu} ,\varphi_\mu\varphi_\nu\rangle
+\langle k,\varphi\varphi\varphi\varphi\rangle$,
where $\varphi$ is a field, $\varphi_\mu$ its derivatives
and $\langle\cdot,\cdot\rangle$ is the duality pairing
between $\Gamma(M,S(JE^*))$ and $\Gamma(M,S(JE))$ induced
by the duality pairing between $JE^*$ and $JE$.
This Hopf algebra is commutative and cocommutative.
Note that the topological properties of this
algebra must be carefully taken into account because
$JE^*$ is an infinite-dimensional Fr\'echet manifold.

In the next section, we shall consider a 
general Hopf algebra bundle $F\overset{\pi}{\rightarrow} M$ whose sections
play the role of Lagrangian densities, where
$F=S(JE^*)$ in Borcherds' case.

\section{The Fock Hopf algebra of classical fields}
Second quantization starts from the construction of an algebra
containing classical fields defined on any number of spacetime points.
The commutative product of this many-body algebra is called
the normal product, and it will be
deformed to define a quantum field algebra.
In Borcherds' paper, the algebra corresponding to the 
normal product of QFT is the symmetric algebra
$S_\bbK(\Gamma(M,F))$ on the space of sections, which is too big to have
a reasonable topology and which is no longer 
geometric, in the sense that 
$S_\bbK(\Gamma(M,F))$ is not the space of sections
of a bundle over a manifold.
This is because this manifold should be
the quotient of $M^n$ by the action of the
symmetric group on $n$ elements, which is generally not a
topological manifold~\cite{Wagner-80}.

To solve that problem, note that for any bundle 
$F\overset{\pi}{\rightarrow} M$
there exists an external tensor product of bundles
$F\boxtimes F \overset{\pi\times \pi}{\longrightarrow} M\times M$
whose space of sections
describes the (completed) tensor product of sections (over $\bbK$),
$\Gamma(M\times M,F\boxtimes F) \cong
\Gamma(M,F) \hat{\otimes}_{\bbK} \Gamma(M,F)$, that is,
$\sigma(x_1,x_2) = \sum \sigma_1(x_1) \otimes \sigma_2(x_2)$.
Moreover, since $\Gamma(M,F)$ is a Hopf algebra
over $C^\infty(M)$, then
$\Gamma(M\times M,F\boxtimes F)$ is a Hopf algebra over
$C^\infty(M^2)$. Similarly,

\begin{dfn}
If $F\overset{\pi}{\rightarrow} M$ is a Hopf algebra bundle,
the normal product of classical fields over
$n$ spacetime points is described by the
\emph{normal product algebra} $\Gamma(M^n,F^{\boxtimes n})$, which
is a Hopf algebra over $C^\infty(M^n)$.
\end{dfn}
Therefore, our normal product is encoded in the tensor product
of sections, corresponding to the external tensor
product of bundles. 
From a physical point of view, if $F=S(JE^*)$ is the bundle
of polynomial Lagrangians of $E$-valued fields, the external
tensor product $\boxtimes$ describes exactly the normal product of
field polynomials at 2 points of $M$: e.g.  the normal
product$\varphi^4(x_1) \partial_\mu \varphi(x_2) \partial^\mu
\varphi(x_2)$
corresponds to the section
$\sigma(x_1,x_2) = \big( (x_1,x_2),\varphi^4 \otimes
\partial_\mu \varphi \partial^\mu \varphi \big)$
of the bundle $F\boxtimes F$ over the point $(x_1,x_2)\in M\times M$.
The exterior tensor product can be performed on any number $n$
of copies of the bundle $F$, giving the Hopf bundle
$F^{\boxtimes n} \overset{\pi^n}{\longrightarrow} M^n$.

To describe QFT, then, we need to define a single algebra which contains
all numbers of points. The difficulty is that the algebras
$\Gamma(M^n,F^{\boxtimes n})$ are defined 
over different rings $C^\infty(M^n)$, one  for each $n$.
It turns out that this problem was solved a long time
ago by Bourbaki.
The first step is to build a ring
$R=\varinjlim C^\infty(M^n)$~\cite{Bourbaki-AlgebraI},
which is the inductive limit of the rings
$C^\infty(M^n)$ corresponding to the map
$\phi_{mn}:C^\infty(M^m)\to C^\infty(M^n)$, with $m \le n$,
defined by
$\phi_{mn}(f)(x_1,\dots,x_n)=f(x_1,\dots,x_m)$.
The inductive limit of algebras over different rings
is also defined by Bourbaki~\cite{Bourbaki-AlgebraI}
and its extension to Hopf algebras is straightforward.
Thus, we obtain a Hopf algebra which is reminiscent of
the Fock space in the sense that it contains any number of points.
\begin{dfn}
If $F\overset{\pi}{\rightarrow} M$ is a Hopf algebra bundle,
the \emph{Fock Hopf algebra} is the inductive limit of
Hopf algebras 
$\HFock=\varinjlim \Gamma(M^n,F^{\boxtimes n})$, which
is a Hopf algebra over the ring 
$\RFock=\varinjlim C^\infty(M^n)$.
\end{dfn}
Note that the Fock Hopf algebra is commutative
iff $F$ is commutative.
The Hopf algebra structure on the Fock algebra is used to perform
its deformation quantization.

We can now wonder whether the Fock Hopf algebra is a space of sections of 
a bundle over some infinite dimensional manifold.
When $M$ can be described by a single
chart to $\bbR^d$, then the answer is yes
and the manifold is $\varprojlim M^n$, which is
a Fr\'echet manifold built on $\varprojlim (\bbR^d)^n$.
If $M$ needs several charts, then the projective limit
topology is not compatible with the structure of a Fr\'echet manifold
and we need more general concepts of infinite-dimensional
manifolds.
We can also wonder whether the definition of 
$\phi_{mn}$ is not too arbitrary. Instead of picking up the
$m$ first points of $(x_1,\dots,x_n)$, we can define an inductive
limit corresponding to any subset of $m$ elements, but by doing so
we recover exactly $\HFock$ and $\RFock$ (because the family of sets
$\{1,\dots,n\}$ is cofinal in the family of subsets of
$\bbN$~\cite{Kothe-I}) so we stick to the
simpler definition because countable inductive limits
have better properties than uncountable ones.

\section{Deformation quantization of $\HFock$}
It remains to quantize the Fock Hopf algebra to recover the
operator product of standard quantum field theory as a special case.
A convenient method to do so is to use 
quantum groups, that Drinfeld created  as a quantization
of algebras~\cite{Drinfeld-86}. His foundation paper
even cites the quantization method of Berezin, Vey, Lichnerowicz, Flato
and Sternheimer (i.e. deformation quantization or star product).
However, the quantization of fields does not use
Drinfeld's quasitriangular structure but its dual,
the \emph{Laplace pairing}, which was
first defined by Lyubashenko~\cite{Lyubashenko-86}.
Rota and Stein called it a Laplace pairing because,
for anticommuting variables, its definition is equivalent
to the Laplace identity of determinants~\cite{RotaStein94}.
Borcherds calls it a \emph{bicharacter}.

\subsection{Laplace pairing}
The problem is now that the Fock Hopf algebra is made of products
of polynomials of smooth sections and their derivatives, whereas
the quantum field amplitudes are distributions.
Therefore, we need to introduce the space 
$\DFock=\varinjlim \calD'(M^n)$, which is the
inductive limit of the spaces of distributions on $M^n$.
The Laplace pairing is 
an $\RFock$-linear map $(\cdot|\cdot): \HFock\otimes_{\RFock}
 \HFock \to \DFock$,
such that, for $a$, $b$ and $c$ in $\HFock$,
$(1|a)=(a|1)=\counit(a)$ and
$(a|bc) = \sum (a\ix1|b) (a\ix2|c)$ and
$(ab|c) = \sum (a|c\ix1) (b|c\ix2)$.
Since the terms $(a\ix1|b) (a\ix2|c)$ and
$(a|c\ix1) (b|c\ix2)$ involve distributions,
the product is only done when wavefront set
conditions are satisfied~\cite{HormanderI}.

In the case of standard quantum field theory,
where $\HFock$ is built from the fiber $F=S(JE^*)$, the Laplace pairing is
determined for $f$ and $g$ in $\Gamma(M,E^*)$ by
$(f\otimes 1|1\otimes g)=\langle f\otimes g,D_+\rangle$, where
$D_+\in \calD'(M^2,E^{{\scriptscriptstyle{\boxtimes}} 2})$
is the Wightman propagator.
This can also be written in a more physical way as
$(\varphi \otimes 1|1\otimes \varphi)=D_+$ or in a non-rigorous
way $(\varphi(x)|\varphi(y))=D_+(x,y)$ in the fiber over $(x,y)$.
This definition is extended to derivatives
of fields by
$(\partial^\alpha \varphi \otimes 1|1\otimes \partial^\beta\varphi)=
\partial^\alpha\partial^\beta D_+$, where
$\alpha$ and $\beta$ are multi-indices.
This pairing is well defined because of the
structural theorem~\cite{Grosser-01}
\begin{eqnarray*}
\calD'(M^2,E^{{\scriptscriptstyle{\boxtimes}} 2}) &=&
\Big(\Gamma_c(M^2,
    (E^*)^{{\scriptscriptstyle{\boxtimes}} 2})\Big)'
\cong  \mathcal{D}'(M^2)\otimes_{C^\infty(M^2)}
     \Gamma(M^2, E^{{\scriptscriptstyle{\boxtimes}} 2})
\\
&\cong & 
\mathcal{L}_{C^\infty(M^2)}
\big(\Gamma(M^2,
    (E^*)^{{\scriptscriptstyle{\boxtimes}} 2}),\mathcal{D}'(M^2)\big).
\end{eqnarray*}

\subsection{Star product}
Quantum group quantization was first defined by Rota and 
Stein~\cite{RotaStein94}, then developed by
Fauser and coworkers~\cite{Fauser,BrouderQG,BrouderMN}.
Its equivalence with the star product was
proved by Hirshfeld~\cite{Hirshfeld}.
Borcherds does not define this product.
\begin{dfn}
Let $F\overset{\pi}{\rightarrow} M$ be a Hopf algebra bundle
and $\HFock$ the corresponding Fock Hopf algebra.
Then, 
$\CFock=\varinjlim \calD'(M^n)\otimes_{C^\infty(M^n)}
\Gamma(M^n,F^{\boxtimes n})$ is a
$\HFock$-Hopf module where the coaction $\beta$ is defined
on $c=u\otimes h$ by
$\beta c=\sum c'\otimes c''=
\sum (u\otimes h\ix1)\otimes h\ix2$.
The \emph{star product}  on $\CFock$  is
defined by
\begin{eqnarray}
c\star d &=& \sum c'd' (c''|d''),
\end{eqnarray}
where $(c''|d'')$ is identified with $(c''|d'')\otimes 1$.
If the Hopf algebra is cocommutative, the
star product is associative.
\end{dfn}
If we consider the example
$c=u\otimes h$ and $d=v\otimes k$ we find
$c\star d=\sum uv (h\ix2|k\ix2) \otimes h\ix1 k\ix1$.
The product $ab (h\ix1|k\ix2)$ is a product
of three distributions which is well-defined 
by the wavefront set condition~\cite{HormanderI}
for standard quantum field theory~\cite{Brunetti2}.
Note that $\CFock$ equipped with the star product is
a sort of generalized Frobenius algebra, in the
sense that $(c\star d|e)=(c|d\star e)$~\cite{BrouderMN}.

For example if $c=(1\potimes 1) \otimes (\varphi\potimes 1)$ and
 $d=(1\potimes 1) \otimes (1\potimes \varphi)$, then
$c\star d=(1\potimes 1) \otimes (\varphi\potimes \varphi)
+ D_+\otimes (1\potimes 1)$ and we recover
 Wick's theorem usually written
$\varphi(x)\star\varphi(y)={:}\varphi(x)\varphi(y){:}+D_+(x,y)$
in QFT textbooks. This completes the quantization of
the Fock Hopf algebra, i.e. the second quantization
of the Hopf algebra bundle $F$.

\subsection{The time-ordered product}
The last step to obtain Green functions of QFT
is to define time-ordered products.
We do this by following the causal approach developed
by Stueckelberg, Bogoliubov, Epstein, Glaser~\cite{Epstein}
and finally Brunetti and Fredenhagen~\cite{Brunetti2}.
Then, the time-ordered product becomes a comodule
morphism $T:\CFock\to \CFock$
and the Wick expansion
of time-ordered products takes the simple form
$T(c)=\sum t(c')c"$, where
$t(c)=(1\otimes \epsilon) (T(c))$ ~\cite{BrouderMN}.
The time-ordered product is defined recursively by
the \emph{causality relation}\footnote{
Borcherds' Gaussian property is a consequence of
the causality relation~\cite{Epstein}.}
saying that $T(cd)=T(c)\star T(d)$ if the 
spacetime support of $c$ is not earlier than
the spacetime support of $d$.
By Stora's lemma\footnote{It can easily be inferred from a remark by
Bergbauer~\cite{Bergbauerdip} that
Stora's lemma only requires a (closed) partial
order on $M$, which is taken to be the causal order
in applications to Lorentzian manifolds.},
the causality
relation and the partial order imply that $T$ is defined
recursively except on the diagonals, where the
distributions have to be extended~\cite{Brunetti2}.
The ambiguity of this extension is organized 
by the renormalization group.

\section{Conclusion}

A second quantization method was described for
any theory whose Lagrangian density is an element of
a cocommutative Hopf algebra bundle. 
Fermions can be taken into account by using
a graded cocommutative Hopf
algebra~\cite{BrouderQG}.
Since we do not require the Hopf algebra to be commutative,
we expect this approach to play a role in the
second quantization of noncommutative geometry.

\end{document}